\documentclass[%
 reprint,
superscriptaddress,
 amsmath,amssymb,
 aps,
]{revtex4-2}

\usepackage{siunitx}
\usepackage{color}
\usepackage{graphicx}
\usepackage{dcolumn}
\usepackage{bm}

\begin{document}

\preprint{APS/123-QED}

\title{Super-Heavy Ions Acceleration Driven by Ultrashort Laser Pulses\\ at Ultrahigh Intensity}

\author{Pengjie Wang}
\affiliation{State Key Laboratory of Nuclear Physics and Technology, School of Physics, CAPT, Peking University, Beijing, China}
\author{Zheng Gong}
\affiliation{State Key Laboratory of Nuclear Physics and Technology, School of Physics, CAPT, Peking University, Beijing, China}
\author{Seong Geun Lee}
\affiliation{Center for Relativistic Laser Science, Institute for Basic Science, Gwangju 61005, Korea}
\affiliation{Department of Physics and Photon Science, Gwangju Institute of Science and Technology, Gwangju 61005, Korea}
\author{Yinren Shou}
\affiliation{State Key Laboratory of Nuclear Physics and Technology, School of Physics, CAPT, Peking University, Beijing, China}
\author{Yixing Geng}
\affiliation{State Key Laboratory of Nuclear Physics and Technology, School of Physics, CAPT, Peking University, Beijing, China}
\author{Cheonha Jeon}
\affiliation{Center for Relativistic Laser Science, Institute for Basic Science, Gwangju 61005, Korea}
\author{I Jong Kim}
\affiliation{Center for Scientific Instrumentation, Korea Basic Science Institute, Daejeon 34133, Korea}
\author{Hwang Woon Lee}
\affiliation{Center for Relativistic Laser Science, Institute for Basic Science, Gwangju 61005, Korea}
\author{Jin Woo Yoon}
\affiliation{Center for Relativistic Laser Science, Institute for Basic Science, Gwangju 61005, Korea}
\affiliation{Advanced Photonics Research Institute, Gwangju Institute of Science and Technology, Gwangju 61005, Korea}
\author{Jae Hee Sung}
\affiliation{Center for Relativistic Laser Science, Institute for Basic Science, Gwangju 61005, Korea}
\affiliation{Advanced Photonics Research Institute, Gwangju Institute of Science and Technology, Gwangju 61005, Korea}
\author{Seong Ku Lee}
\affiliation{Center for Relativistic Laser Science, Institute for Basic Science, Gwangju 61005, Korea}
\affiliation{Advanced Photonics Research Institute, Gwangju Institute of Science and Technology, Gwangju 61005, Korea}
\author{Defeng Kong}
\affiliation{State Key Laboratory of Nuclear Physics and Technology, School of Physics, CAPT, Peking University, Beijing, China}
\author{Jianbo Liu}
\affiliation{State Key Laboratory of Nuclear Physics and Technology, School of Physics, CAPT, Peking University, Beijing, China}
\author{Zhusong Mei}
\affiliation{State Key Laboratory of Nuclear Physics and Technology, School of Physics, CAPT, Peking University, Beijing, China}
\author{Zhengxuan Cao}
\affiliation{State Key Laboratory of Nuclear Physics and Technology, School of Physics, CAPT, Peking University, Beijing, China}
\author{Zhuo Pan}
\affiliation{State Key Laboratory of Nuclear Physics and Technology, School of Physics, CAPT, Peking University, Beijing, China}
\author{Il Woo Choi}
\email{iwchoi@gist.ac.kr}
\affiliation{Center for Relativistic Laser Science, Institute for Basic Science, Gwangju 61005, Korea}
\affiliation{Advanced Photonics Research Institute, Gwangju Institute of Science and Technology, Gwangju 61005, Korea}
\author{Xueqing Yan}
\email{x.yan@pku.edu.cn}
\affiliation{State Key Laboratory of Nuclear Physics and Technology, School of Physics, CAPT, Peking University, Beijing, China}
\author{Chang Hee Nam}
\email{chnam@gist.ac.kr}
\affiliation{Center for Relativistic Laser Science, Institute for Basic Science, Gwangju 61005, Korea}
\affiliation{Department of Physics and Photon Science, Gwangju Institute of Science and Technology, Gwangju 61005, Korea}
\author{Wenjun Ma}
\thanks{wenjun.ma@pku.edu.cn}
\affiliation{State Key Laboratory of Nuclear Physics and Technology, School of Physics, CAPT, Peking University, Beijing, China}






\begin{abstract}
The acceleration of super-heavy ions (SHIs, mass number $ \sim $ 200) from plasmas driven by ultrashort (tens of femtoseconds) laser pulses is a challenging topic waiting for breakthrough. The detecting and controlling of the ionization process, and the adoption of the optimal acceleration scheme are crucial for the generation of highly energetic SHIs. Here, we report the experimental results on the generation of deeply ionized super-heavy ions (Au) with unprecedented energy of 1.2 GeV utilizing ultrathin targets and ultrashort laser pulses at the intensity of $ 10^{22}\ {\rm W}/{\rm cm}^{2} $. A novel self-calibrated diagnostic method was developed to acquire the absolute energy spectra and charge state distributions of Au ions abundant at the charge state of 51+ and extending to 61+. The measured charge state distributions supported by 2D particle-in-cell simulations serves as an additional tool to inspect the ionization dynamics associated with SHIs acceleration, revealing that the laser intensity is the crucial parameter over the pulse duration for Au acceleration. Achieving a long acceleration time without sacrificing the strength of acceleration field by utilizing composite targets can substantially increase the maximum energy of Au ions.
\end{abstract}

\maketitle


\section{introduction}
Laser-driven acceleration of energetic ions is an attractive topic\cite{Dover2020prl, Braenzel2015prl, Willingale2006prl} owing to its unique features such as ultrahigh accelerating gradient, micrometer-scale source size, high beam density and low emittance\cite{Nishiuchi2015pop, Domanski2018lpb, Li2019njp}. In particular, laser-driven ion acceleration promises to generate super-heavy ion beams with ultrahigh intensity and multiple charge states. Such an super-heavy ion source is highly desired for numerous applications including fission-fusion reaction\cite{Habs2011apb} driven by ultrahigh-intensity thorium beams, heavy ion accelerators\cite{Busold2015sr, Ostroumov2000prab} that require the injection of multiple-charge-state beams, heavy-ion fusion research\cite{Kawata2016mre} and the generation of warm dense matter\cite{Bang2015sr}. Compared with the acceleration of low-Z ions (80 MeV/nucleon for C) and mid-Z ions (10-20 MeV/nucleon for $ {\rm Al}^{13} $, $ {\rm Fe}^{26} $ and $ {\rm Ag}^{47} $)\cite{Nishiuchi2015pop, Jung2013pop, Palaniyappan2015nc, Nishiuchi2020prr, Ma2019prl}, the accessible energy of super-heavy ions (SHIs) from laser-plasma interactions is quite low for a long time. 2 MeV/nucleon of $ {\rm Pb}^{82} $\cite{Clark2000prl} and 5 MeV/nucleon of $ {\rm Au}^{79} $\cite{Lindner2019ppcf} ions were reported by employing 100s J long-pulse lasers, which is unfavorable for applications that need high-repetition rate driven by compact and economic laser systems.

\begin{figure}[b]
	\includegraphics{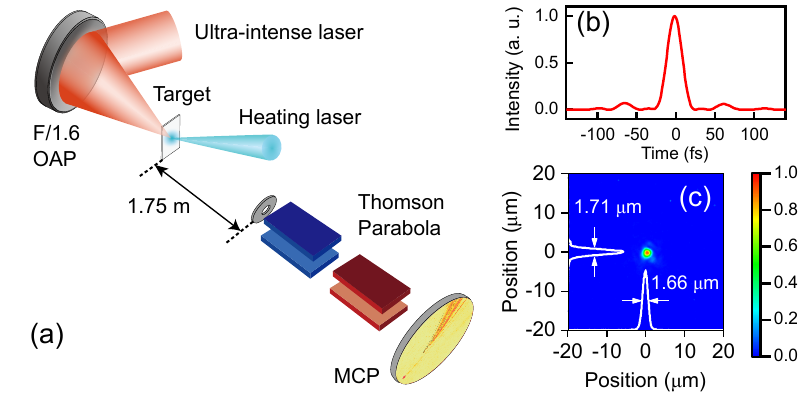}
	\caption{\label{fig:setup} (a) Experimental setup. The temporal profile (b) and the focal spot (c) of driving laser pulses.}
\end{figure}

Generally, the acceleration of SHIs would be suppressed by the inevitable contamination attached to the surface of solid targets, where the undesired hydrocarbons are more readily accelerated due to their higher charge-to-mass ratio. Resistive heating\cite{Hegelich2006nature} or laser heating\cite{Safronov2018pop} have been proven as useful methods to remove the contamination and thus facilitate heavy ion acceleration. However, SHIs are still hard to be efficiently accelerated, because they need to be ionized to high charge states as early as possible to experience a long acceleration time\cite{Petrov2016pop}. The accessible intensity of $ 10^{20}\ {\rm W}/{\rm cm}^{2} $ for picosecond laser pulses that have been widely utilized for the acceleration of low-Z and mid-Z ions, is insufficient to collisionless deeply ionize SHIs. The deep ionization of SHIs relies on the collisional ionization in thermalized plasma, which requires higher laser energy but results in a lower scaling factor of the maximum energy of SHIs with laser energy\cite{Petrov2017ppcf}. On the other hand, the state-of-art multi-petawatt femtosecond lasers have recently been capable of delivering an intensity higher than $ 10^{22}\ {\rm W}/{\rm cm}^{2} $. When such an ultraintense femtosecond laser pulse irradiates an ultrathin target\cite{Domanski2018lpb}, huge ponderomotive force can pile up bulk electrons to build up an ultrastrong charge separation field, resulting in deep collisionless ionization\cite{Kawahito2020pop} and prompt injection\cite{Li2019sr}. It has been predicted that 10s GeV thorium ($ ^{232} $Th) ions with very high charge states can be obtained\cite{Domanski2018pla} by shooting an ultrathin target at the intensity of $ 10^{23}\ {\rm W}/{\rm cm}^{2} $ with the upcoming 10 petawatt and exawatt class lasers\cite{Danson2019hpl}. However, the proof-of-principle experiment on highly charged super-heavy ion acceleration, driven by the ultrashort femtosecond laser pulses at such ultrahigh intensity, had not yet been demonstrated so far.

Here, we experimentally and numerically study the generation of deeply ionized energetic Au ions by using ultrashort femtosecond laser pulses at the intensity of $ 10^{22}\ {\rm W}/{\rm cm}^{2} $. The energy spectra and the absolute charge state distributions of Au ions were measured by a novel self-calibrated diagnostic method. The charge state distributions from single- and double-layer targets provide additional information to inspect the ionization and acceleration process. With the help of 2D particle-in-cell (PIC) simulations, the influence of the ionization dynamics on the acceleration process is discussed.

\section{experimental setup and targets}

The experiments were carried out in the 4 petawatt Ti: sapphire laser facility\cite{Sung2017ol} located at the Center for Relativistic Laser Science (CoReLS). The p-polarized laser pulses were tightly focused onto targets by an f/1.6 off-axis parabolic mirror with the energy of 14-15 J as shown in Fig.~\ref{fig:setup}(a). The temporal profile of the pulse was measured with a SPIDER right after the pulse compressor. The full-width-at-half-maximum (FWHM) duration was 22 fs as shown in Fig.~\ref{fig:setup}(b). The focal spot characterization was performed with a pair of lenses and a 12-bit CMOS camera in the 100 TW mode by attenuating the laser energy with partial reflection mirrors\cite{Yoon2019oe}. Figure~\ref{fig:setup}(c) shows the measured focal spot. The best focal spot size measured in the campaign had a near-diffraction-limited size of $ 1.71\times1.66 $ $ \mu {\rm m}^{2} $ FWHM, and 32$ \% $ of the laser energy was concentrated in the FWHM area.

Calculated from the pulse duration and focal spot measurement, and taking into account of the fluctuation in the focal spot quality, the on-target intensity was $ 1.1\ \pm\ 0.4\ \times\ 10^{22}\ {\rm W}/{\rm cm}^{2} $, which corresponds to a normalized laser amplitude of $ a_{0} $ = $ eE_{0}/mc\omega \simeq$ 57-84, where $ e $, $ m $, $ c $, $\omega$, and $ E_{0} $ are electron charge, mass, light speed, laser frequency, and electric field amplitude, respectively. The contrast ratio of the laser pulse was better than $ 10^{12} $ up to 2 ps before the main pulse by employing a double plasma mirror system\cite{Choi2020ol}, which avoids the prepulse-heating and premature expansion of the targets happening for low-contrast laser\cite{Nishiuchi2020prr}.

Single-layer ultrathin foil targets made of Au, Ag, Cu, and diamond-like carbon, and double-layer targets composed of carbon nanotube foams (CNF) and ultrathin Au foils, were used in the experiments. In contrast to previous work, a water-cooling chemical vapor deposition system designed and developed at Peking University enables the online deposition of CNF onto ultrathin Au foils, which can prevent the melting of ultrathin Au foils. The density of the CNF is $ 2.3\ \pm\ 0.5\ {\rm mg/cm^{3}} $, corresponding to electron density of $ 0.4\ \pm\ 0.1\ n_{c} $ when the atoms are fully ionized, where $ n_{c} = m\omega^{2}\varepsilon_{0}/e^{2} $ is the critical density of the plasma. The laser pulse was focused onto the CNF side with an incident angle of $2.5^{\circ}$.

\begin{figure}[b]
	\includegraphics{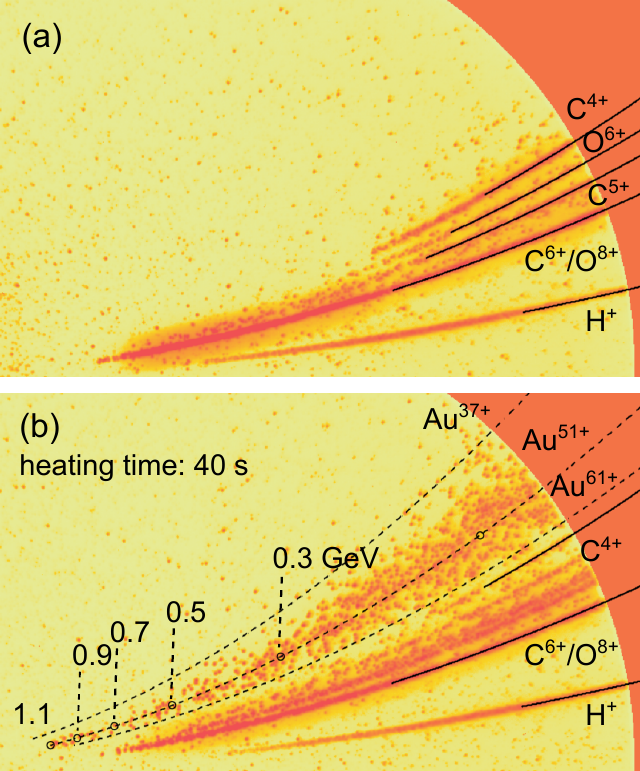}
	\caption{\label{fig:rawdata} Raw data measured with TPS from (a) a 20-nm diamond-like carbon target and a double-layer Au target (60 $ \mu {\rm m} $ CNF + 150 nm Au).}
\end{figure}

\begin{figure*}
	\includegraphics{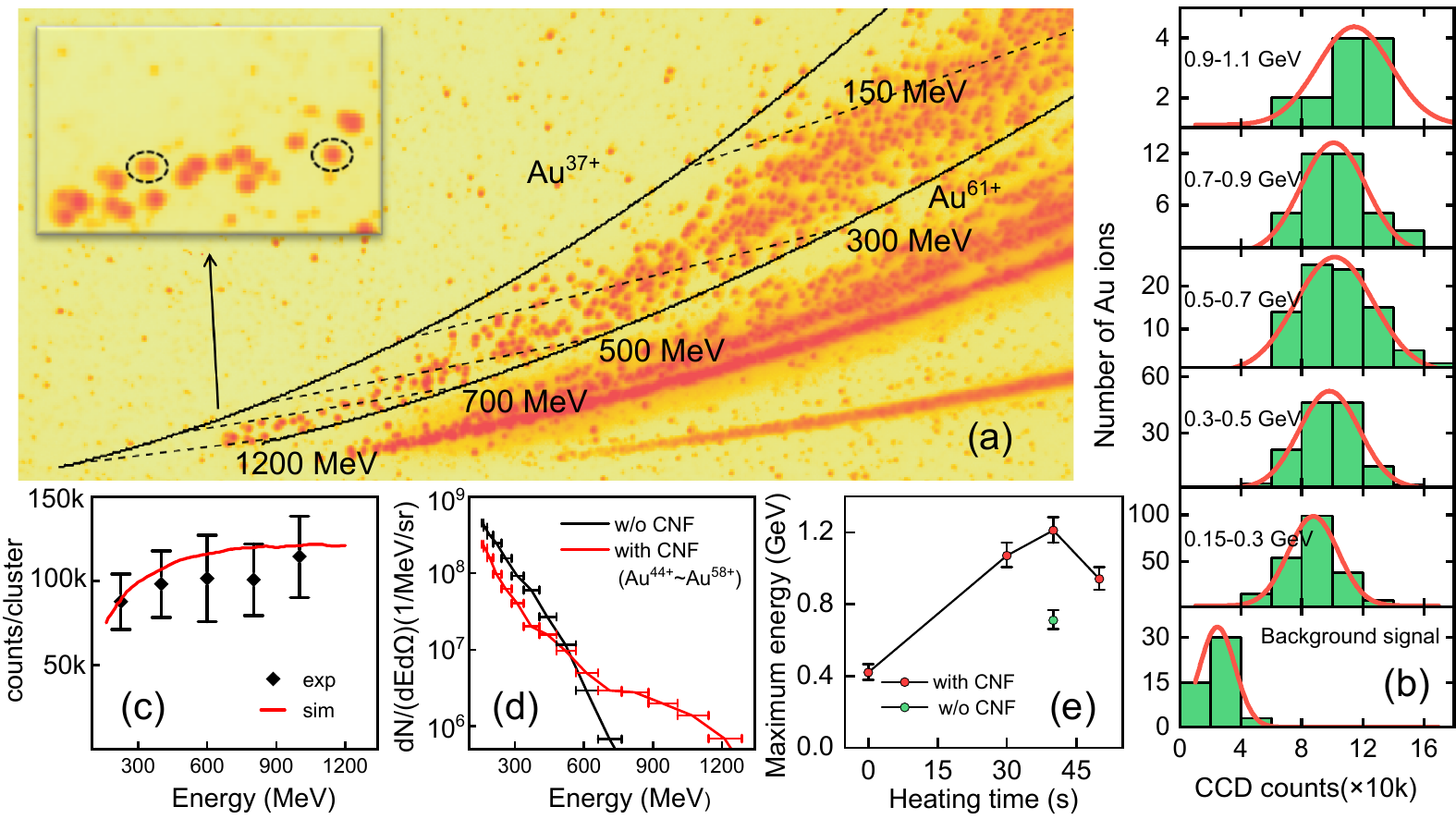}
	\caption{\label{fig:sie and spectra}(a) Schematic of the statistics of single-ion events: the solid lines represent the parabolas of $ {\rm Au}^{37+} $ and $ {\rm Au}^{61+} $, and the dashed lines represent the constant energy lines. The inset shows the zoomed clusters of ion signals. (b) Statistical results of single-ion events. The red lines are the Gaussian fitting curves. (c) Responses for Au ions. (d) Energy spectra of Au ions. (e) Maximum energy of Au ions without heating and with varying heating time.}
\end{figure*}

A continuous wave diode laser with the maximum power of 500 mW was utilized to heat the rear surface of targets to remove the contamination layer before the main laser irradiation. Ions were detected by a Thomson parabola spectrometer (TPS) equipped with a microchannel plate (MCP) with a phosphor. Ions hitting on the MCP assembly will produce optical signals which are imaged by a 16-bit charge-coupled device (CCD) camera. In order to acquire the charge state distribution of Au ions, the TPS was placed 1.75 m away from the targets, and the diameter of the collimating tungsten pinhole is 310 $\mu m$, which issues in an ultra-small detection solid angle of $ 2.5\times10^{-8}\ {\rm sr} $. Only a few hundreds of Au ions can pass the pinhole and arrive at the MCP per shot. A maximum electric field of 18.2 kV/cm was applied in TPS to disperse the Au ion Thomson parabolas as much as possible on the MCP to produce single-ion events.

\section{experimental results and analysis}
\subsection{Experimental results}

\begin{figure}[t]
	\includegraphics{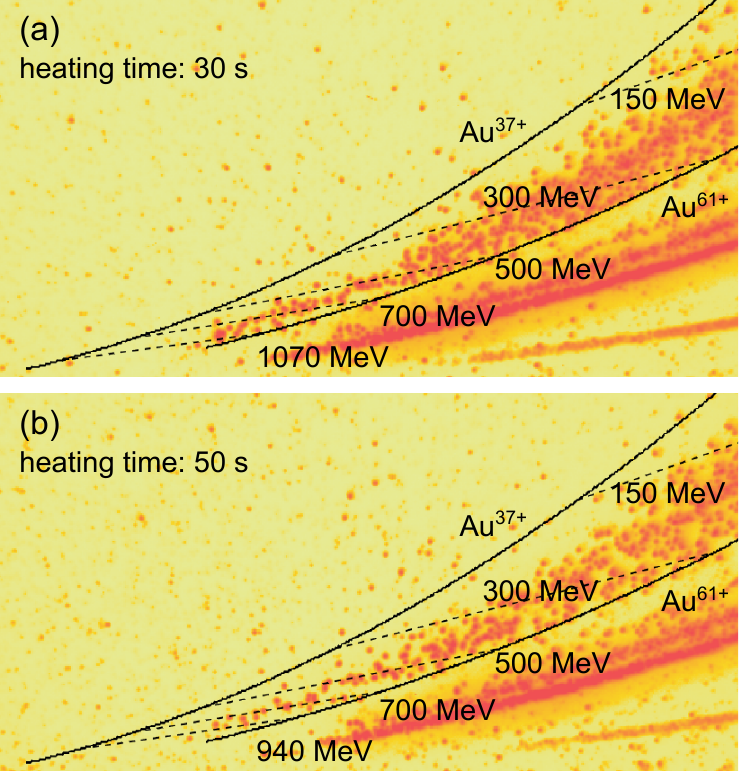}
	\caption{\label{fig:rawdata_more} Raw data measured with TPS from double-layer Au targets (60 $ \mu {\rm m} $ CNF + 150 nm Au) varyring heating time of 30 s (a) and 50 s (b).}
\end{figure}

Figures~\ref{fig:rawdata}(a) and~\ref{fig:rawdata}(b) show the raw data, measured with TPS, for a 20-nm diamond-like carbon and a double-layer Au target (60 $ {\rm \mu m} $ CNF\ +\ 150\ nm Au), respectively. The TPS-MCP assembly is a common diagnostic tool in laser-driven ion acceleration experiments. Electric and magnetic fields in TPS can disperse ions in two orthogonal directions, based on charge-to-mass ratio and energy of ions. Ions with different charge-to-mass ratios form different parabola-like traces on the surface of MCP. For the diamond-like carbon target, the traces of $ {\rm H^{+}} $, $ {\rm C^{6+}} $/$ {\rm O^{8+}} $, $ {\rm C^{5+}} $, $ {\rm O^{6+}} $, and $ {\rm C^{4+}} $ can be clearly distinguished. C and O ions with charge-to-mass ratio $<$ 1/3 are not observed, which indicates the acceleration of $ {\rm C^{<4+}} $ and $ {\rm O^{<6+}} $ are insignificant at this intensity. Based on this observation, it is inferred that the cluster signals in the region of charge-to-mass ratio $<$ 1/3 for double-layer Au targets are caused only by the Au ions. As shown in Fig.~\ref{fig:sie and spectra}(a), the clusters have similar shapes and clear boundaries, indicating they come from single-ion events. We summed the CCD counts for each distinct cluster as the response of a single ion, and did the statistics as a function of the ion energy. The histogram of the responses collected in the region of charge-to-mass ratio $<$ 1/3 from multiple shots is shown in Fig.~\ref{fig:sie and spectra}(b). The responses demonstrate a clear dependence on the energy of Au ions, as depicted in Fig.~\ref{fig:sie and spectra}(c). We performed a series of calculations to simulate the MCP response curve considering the energy, incident angle, and incident positions of the ions (see part B in this section). The simulations agree well with the experimental results, confirming each cluster as the signal from a single Au ion. At the bottom of Fig.~\ref{fig:sie and spectra}(b), the responses of clusters outside of the Au ion region are also presented, as the background signal. They might come from other scattered radiations hitting on the MCP, of which counts are significantly lower than that of the Au ions.

\begin{figure}[b]
	\includegraphics{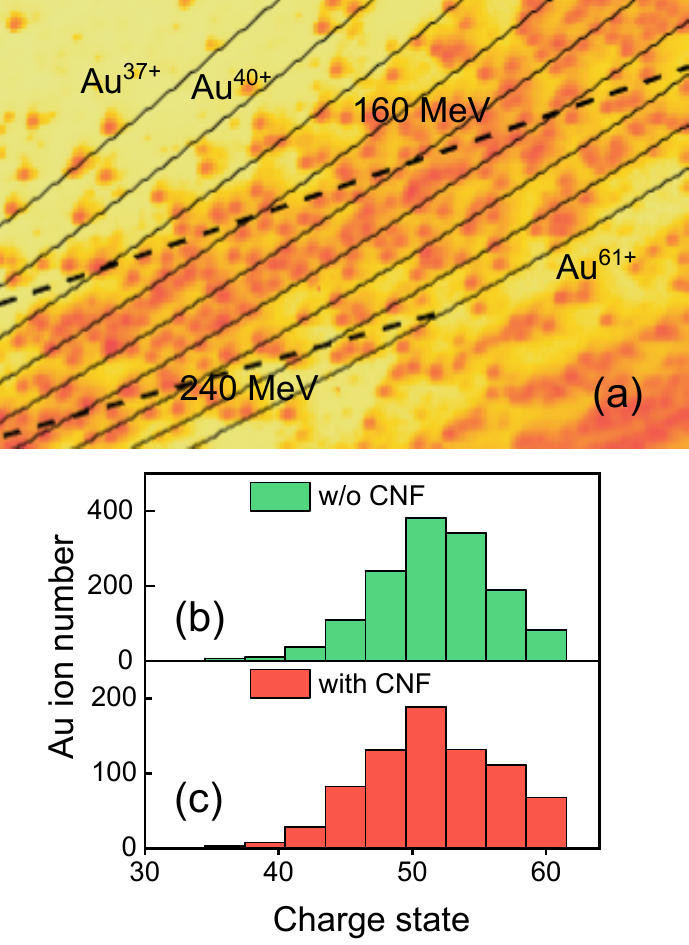}
	\caption{\label{fig:csd} (a) Schematic of the statistics of the charge state distributions: solid lines represent the constant charge-state lines. Charge state distributions of Au ions from a single-layer (b) and a double-layer (c) targets.}
\end{figure}

Based on the obtained calibration, the absolute energy spectra and charge state distributions of Au ions can be extracted. Figure~\ref{fig:sie and spectra}(d) shows two representative energy spectra. Note that our self-calibrated method can be further confirmed, by comparing the order of magnitude from our results with that using CR-39 detectors\cite{Lindner2019ppcf}. The maximum Au energy from 150-nm single-layer targets is 710 MeV (3.6 MeV/nucleon), 3.6 times higher than the previous results obtained from 14 nm Au foils irradiated by femtosecond lasers\cite{Braenzel2015prl}. For the double-layer target, the maximum Au energy is 1.2 GeV (6.1 MeV/nucleon) at optimal CNF thickness of 60 $ {\rm \mu m} $, enhanced by 1.7 times compared to single-layer target. The integrated number of Au ions with energy$>$200 MeV is $\rm 1.2\times10^{10}\ sr^{-1}$. We found the heating time of the targets significantly influences the energy spectra of Au ions, which is in agreement with previous studies. Figure~\ref{fig:sie and spectra}(e) shows the statistics of the maximum Au energy as a function of the heating time. It can be seen that, without heating, the maximum energy of Au ions is lower than 440 MeV. By varying the heating time from 30s $ \sim $ 50s, the maximum energy of Au ions from single-layer and double-layer targets are in the range of 660-710 MeV and 930-1200 MeV, respectively. The corresponding raw data of varying heating time are shown in Figs.~\ref{fig:rawdata_more}(a) and ~\ref{fig:rawdata_more}(b).

The absolute charge state distributions were obtained by counting the Au ions in striped areas segmented by the constant charge-state lines shown in Fig.~\ref{fig:csd}(a). Technically, all CCD counts in one striped area, were summed after subtracting the background, and then were divided by the calibrated response of Au ions to obtain the ion numbers in that area. Limited by the resolution of TPS, the statistic division of charge states is set to be 3 in Figs.~\ref{fig:csd}(b) and~\ref{fig:csd}(c) for a clear observation of the trend. It is found that there are no significant differences between the cases of single-layer and double-layer targets, which will be analyzed below with the help of numerical simulations. By carefully examining the data, we conclude that highly charged state of up to 61+ (Ar-like Au ion) was obtained, and the abundant state is 51+ (Ni-like Au ion).

\subsection{Simulations confirming the self-calibrated\\ ion responses}

The calibration of MCP for Au ions in TPS-MCP assembly has not been reported in literature. The single-ion events, measured here, of the clusters can give a natural self-calibration. In order to confirm such self-calibrations, we simulated response curves for Au, Ag, Cu ions with different energy, and compared them with the measured data in the experiments. 

\begin{figure}[b]
	\includegraphics{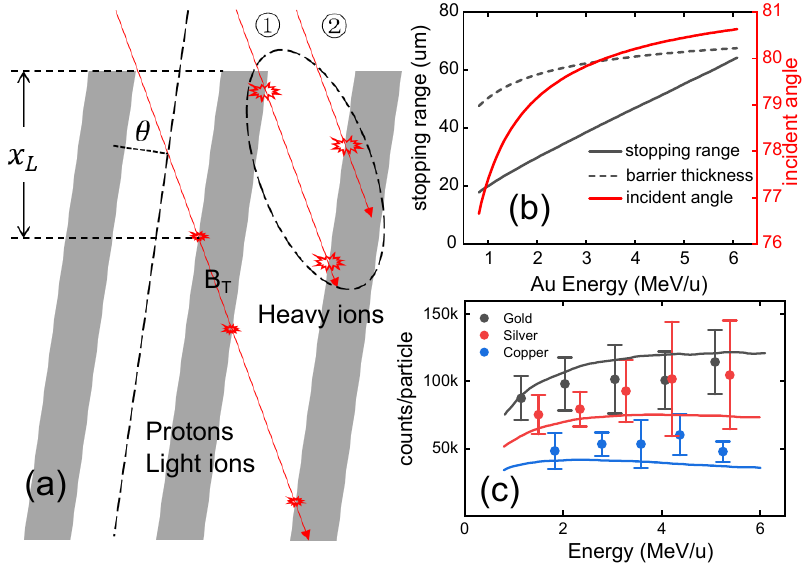}
	\caption{\label{fig:mcp_simulation} (a) Schematic of ions interacting with MCP. The bomb symbols show the generation of secondary electrons on surface of MCP channels. (b) The stopping range (black solid line) of Au ions in MCP materials and the thickness of barrier (black dashed line) needed to pass through to enter into another channel, depicted as $ {\rm B}_{\rm T} $. The incident angle $ \theta $ (red solid line). (c) Responses of heavy ions obtained from experiments (solid dots) and from calculations (solid lines).}
\end{figure}

A MCP consists of millions of micro channels made of 10s-um-thin conductive glass capillaries. Once an ion strikes the inner wall of a channel, multiple secondary electrons are emitted into the channel. Those electrons are accelerated by electrical field applied along MCP, and may strike the opposite wall in this channel inducing further electrons repeatedly. Finally, the detected signal can be amplified exponentially. The response of MCP, for protons and light ions (C, O), has been calibrated before\cite{Jeong2016rsi, McIlvenny2019ji, Prasad2013rsi, Prasad2010nima}.

The theoretical model considered in Refs\cite{Jeong2016rsi, Prasad2010nima} was used to analyze the MCP response for Au ions.

\begin{equation}
	R_{channel}\propto\frac{1}{\cos\theta}\left(\frac{dE}{dx}\right)\Delta xg,
	\label{subeq:1}
\end{equation}

where $ \theta $ is the incident angle of ions in respect to the surface of MCP channel and $(dE/dx)$ is the electronic stopping power of ions striking the surface of MCP channel one time. $ \Delta x $ is the depth above where secondary electrons induced by ions can enter into this channel\cite{Then1990jns}. $ g $ is the gain in this channel, which can be calculated as $ g=e^{k(L-x_{L})/L} $. $ k $ is a constant, and $ L $ is the thickness of MCP. $ x_{L} $ is the depth from the surface of MCP, where ions hit onto inner surface of the channel. These parameters are marked on Fig.~\ref{fig:mcp_simulation}(a). Here we used the MCP (F2226-14P130) from Hamamatsu, with thickness ($ L $) of 1 mm and bias angle of $ 8^{\circ} $.

For protons with energy $ > $ 10 MeV, the stopping range is larger than the thickness of the wall of the channel. In this case, one proton can pass through more than one channel to generate more electrons. The final response needs to sum up all responses from all the channels. In our experiment, the energy of Au ions is lower than 7 MeV/nucleon. Figure~\ref{fig:mcp_simulation}(b) shows the stopping range and incident angle of an Au ion into the MCP materials. For 6 MeV/nucleon Au ions, the maximum stopping range is 64 $ \mu {\rm m} $, which remains smaller than the barrier thickness of the wall along the ion path $ {\rm B}_{\rm T} = 68\ \mu {\rm m} $. As a result, most of the Au ions only induce signals from one channel. Nevertheless, there is still a chance that one ion can induce signals from 2 channels, as we show in the case 1 ($ Response_{1} $) in Fig.~\ref{fig:mcp_simulation}(a). In case 2 ($ Response_{2} $), Au ions hit into the channel of MCP directly, only generating one group of secondary electrons. Therefore, the final response can be written as:

\begin{equation}
	Response=\sum_{i=1}^{2}P_{i}\cdot Response_{i}\propto\frac{1}{\cos\theta}\left(\frac{dE}{dx}\right)_{e}\Delta x\bar{g},
	\label{subeq:2}
\end{equation}

\begin{figure*}
	\includegraphics{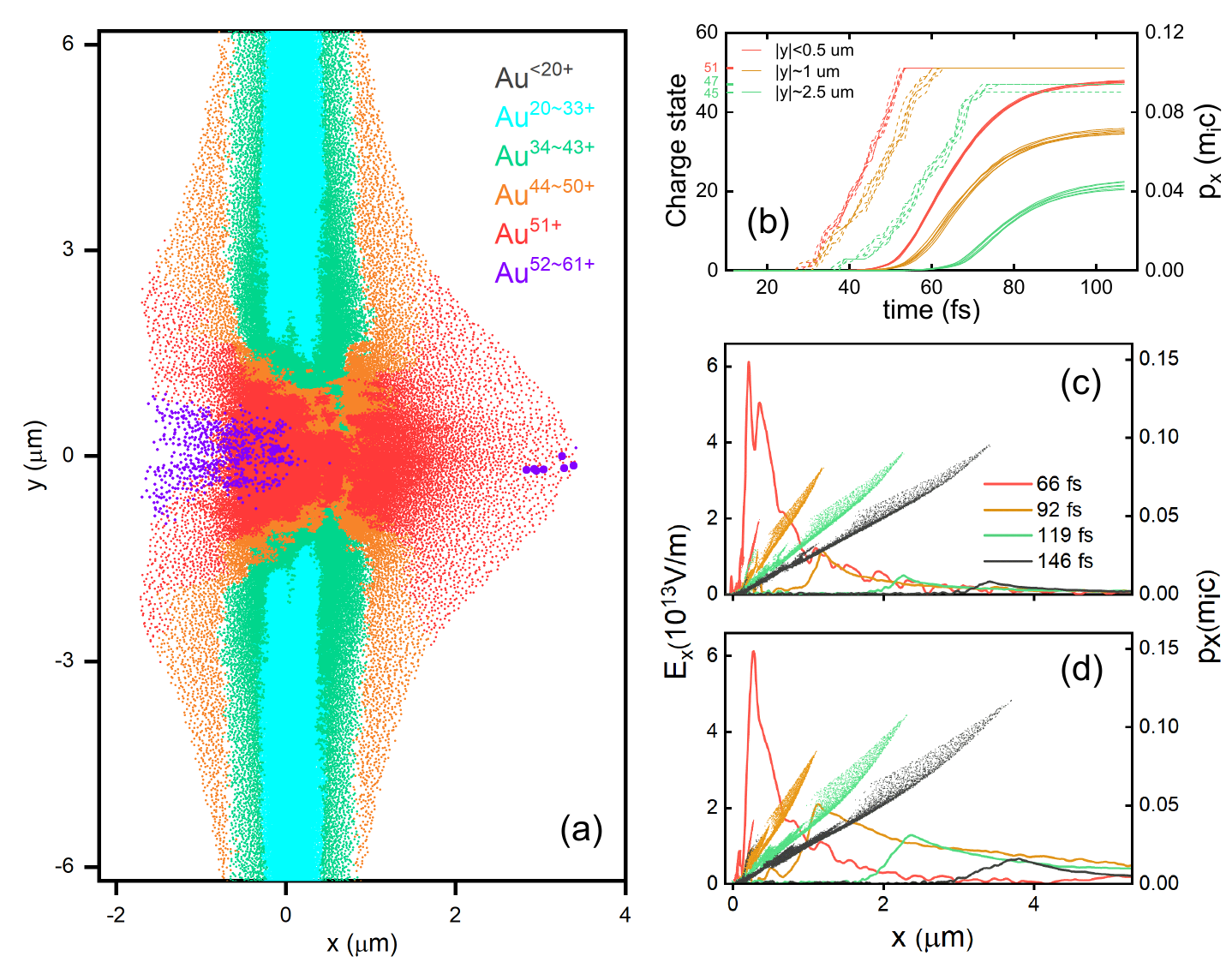}
	\caption{\label{fig:pic_acceleration}(a) Spatial distribution of Au ions from a 150-nm single layer Au target for $ a_{0} $ = 60 at t = 146 fs. (b) Evolution of the charge states (dashed lines) and the longitudinal momentum $ p_{x} $ (solid lines) of Au ions originating from different transverse positions $|{\rm y}|$ for the target in (a): each line represents one macro-particle. Evolution of the acceleration field $ E_{x} $ (solid lines) and $ x-p_{x} $ phase space (scattered dots) for the single-layer target (c) and the double-layer target (d).}
\end{figure*}

The incident angle $ \theta $ can be calculated based on the energy and charge state of Au ions, the parameters of TPS and bias angle of MCP. $ (dE/dx)_{e} $ can be extracted from SRIM simulations\cite{Ziegler2010nimb}. $ \bar{g} $ can be obtained using $ \bar{x}_{L} $, the average ion penetration depth inside the channel. Moreover, we also did the experimentally statistics based on single-ion events, and calculations based on our model, of responses for Ag and Cu ions within similar energy range. The solid dots in Fig.~\ref{fig:mcp_simulation}(c) shows the responses of Au, Ag, and Cu ions with respect to their energies obtained in the experiments and the solid lines for calculations. According to Bethe-Bloch equation, the electronic stopping power of ions traversing a matter is proportional to the square of charge state of the incident ion. Therefore, the response for Au ions would be the largest among these three kinds of ions because of their highest energy deposition in MCP. The theoretical model fits well with the experimental results for three kinds of heavy ions.

\section{PARTICLE-IN-CELL SIMULATIONS AND DISCUSSION}

We performed a series of 2D PIC simulations to illustrate the ionization and acceleration process utilizing the EPOCH code\cite{Arber2015ppcf}. The simulation region with a size of $ 90 \lambda_{0}\times30 \lambda_{0} $ is uniformly divided into 9000 $\times$ 3000 mesh grids. A linearly polarized laser with wavelength $ \lambda_{0} = 0.8\ \mu {\rm m} $ is incident from the left boundary with FWHM diameter of 1.69 $ \mu m $ and FWHM duration of 20 fs. A uniform 150-nm Au target with an atom density of $ n_{a} = 5.96\times10^{28}/{\rm m}^{3} $ is placed at $ 0<x<150\ {\rm nm} $. In the case of the double-layer target, a near-critical-density plasma slab (representing the CNF) with electron density of $ 0.4\ n_{c} $ is placed ahead of the Au target. Here, both carbon and Au targets are initialized as neutral atoms\cite{Yu2018ppcf} which are represented by 20 (Carbon) and 50 (Au) macro particles per cell, respectively. The atoms are gradually ionized by the incident laser pulses. The field ionization model adopts the ADK ionization rate\cite{Ammosov1987spj} where the equation is averaged over all possible magnetic quantum numbers. The corrected effect from barrier-suppression ionization is incorporated into ADK ionization rate as well\cite{Posthumus2004rpp}. Collisional ionization is not included in the simulations due to reasons given below.

We firstly investigate the influence of the ionization dynamics on Au acceleration in the single-layer target. The spatial distribution of Au ions at t = 146 fs irradiated by a laser pulse of $ a_{0} $ = 60 is shown in Fig.~\ref{fig:pic_acceleration}(a). At this moment, the intensity peak of the laser pulse has been reflected for 87 fs from t = 59 fs, and the primary acceleration process is completed. Ions near the center of the focal spot have higher charge states than those away from the center, as a consequence of the intensity gradient of the driving laser. Judging from their $ x $ positions, one can find that $ {\rm Au}^{\geqslant51+} $ are more energetic than $ {\rm Au}^{<51+} $. They are ionized in the strongest sheath field and accelerated with higher field gradients due to their higher charge-to-mass ratio. 

The evolution of the charge states and the longitudinal momentum of Au ions, grouped according to their transverse positions, are examined, as shown in Fig.~\ref{fig:pic_acceleration}(b). Ions at the center of the focus ($|{\rm y}| < 0.5\ \mu {\rm m}$, red lines) are firstly ionized to $ {\rm Au}^{51+} $ with the maximum ionization rate, and then undergo the longest acceleration time starting from t = 45 fs. Ions near the edge of the focal spot ($|{\rm y}| \simeq 1.5\ \mu {\rm m} $, yellow lines) are ionized to $ {\rm Au}^{51+} $ 10 fs later, and their final energy is significantly lower due to the shorter acceleration time. If ions originate further away from the center of the focus ($|{\rm y}| \sim 2.5\ \mu {\rm m}$, green lines), they start to be ionized and reach their final charge states (45-47) even later, which leads to a more inefficient acceleration as compared to those from the center of the focal spot. 

The PIC simulations indicate that the evolution of the acceleration field plays a crucial role in the acceleration of SHIs. As seen from the temporal evolution of $ E_{x} $ at y = 0 for the single-layer and double-layer targets, in Figs.~\ref{fig:pic_acceleration}(c) and~\ref{fig:pic_acceleration}(d), respectively, the maximum strengths of $ E_{x} $ for the two cases are similar, but the $ E_{x} $ of the double-layer target decays much more slowly. The distributions of Au ions in $ x-p_{x} $ phase space are depicted in Figs.~\ref{fig:pic_acceleration}(c) and~\ref{fig:pic_acceleration}(d) as well. At t = 66 fs, when the laser peak has been reflected while the majority of the super-ponderomotive electrons\cite{Ma2019prl} have not arrived at the second-layer, $ E_{x} $ acting on Au ions are almost the same for the two cases. However, at t = 92, 119, 146 fs, after the arrival of the super-ponderomotive electrons, $ E_{x} $ exerted on the most energetic ions for the double-layer target are the twice of that for the single-layer target. Before $ E_{x} $ becomes too weak for acceleration, the most energetic Au ions in double-layer targets virtually experience a longer acceleration time in stronger fields, which eventually results in their higher energy. 

\begin{figure}[t]
	\includegraphics{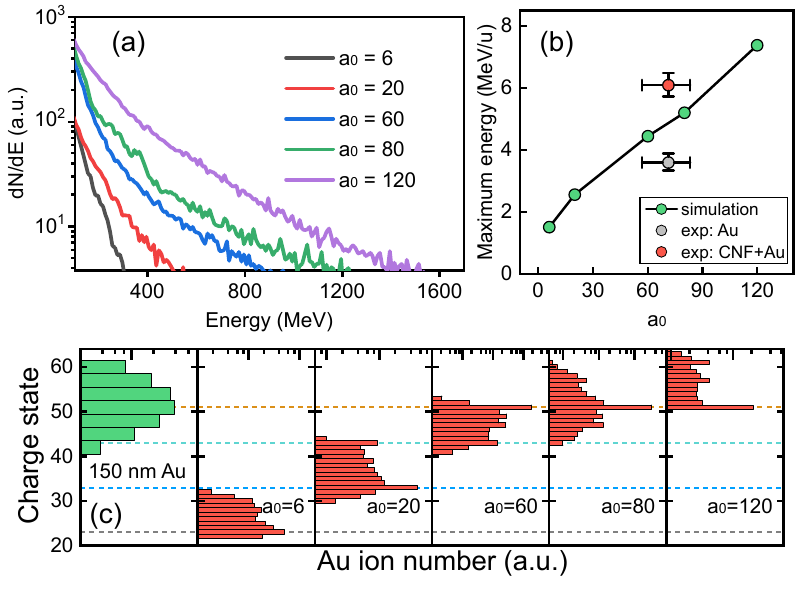}
	\caption{\label{fig:pic_csd} (a) Energy spectra of Au ions at different intensity controlled by changing the pulse duration while keeping the laser energy constant. (b) Dependence of the maximum Au ion energy on laser intensity. Symbols with error bars mark the experimental results. (c) The charge distributions of Au ions in experiments (green bars) and PIC simulations (red bars).}
\end{figure}

Instead of employing the double-layer targets, the acceleration time can be prolonged by experimentally stretching the driving pulses as well, which nevertheless would lead to the reduction of laser intensity for a given laser energy. For low-Z ions, it was found that the optimal pulse duration is about 100-150 fs for a given laser as the tradeoff between the acceleration time and acceleration gradient\cite{Schreiber2006prl}. To figure out the dominant factor between duration and intensity for SHIs acceleration, we carried out a series of simulations by varying the pulse duration from 5 fs to 2 ps while keeping the laser energy constant. Figure~\ref{fig:pic_csd}(a) shows the energy spectra from 150 nm Au targets at different laser intensity. The dependence of the maximum energy of Au ions on the intensity is presented in Fig.~\ref{fig:pic_csd}(b). The results indicate that the laser intensity is more crucial for the acceleration of Au ions over the pulse duration, even when the pulse duration is as short as 5 fs in the case of $ a_{0} $ = 120, which is very different from the result of low-Z ions. The use of double-layer targets leads to the prolongation of the acceleration time without sacrificing the acceleration field strength, which is highly favorable for SHIs acceleration.

In the simulations, we find that the charge state distributions of Au ions are closely related to the laser intensity. Figure~\ref{fig:pic_csd}(c) exhibits the charge state distributions of Au ions obtained in the simulations (red bars) at different laser intensities for a 150-nm Au target. The statistics include ions with energy higher than 10$ \% $ of the maximum energy and within 10 degrees (half-angle) from the target normal. With the increase of the intensity from $ a_{0} $ = 6, 20 to 60, the peak of charge states varies among $ {\rm Au}^{23+} $, $ {\rm Au}^{33+} $, and $ {\rm Au}^{51+} $, respectively. Such abundant concentration is due to the leap of the ionization energy among the the sequential charge states. By comparing the experimental charge state distributions with simulation ones, an estimation of acceleration field strength, and furthermore, the on-target laser intensity, can be performed\cite{Ciappina2019pra}. Figure~\ref{fig:pic_csd}(c) shows that the experimental charge state distributions elucidate an intermediate state between the simulated results of $ a_{0} $ = 60 and $ a_{0} $ = 80, which is in good agreement with the laser intensity employed in this experiment.

Based on the experimental result of the similar charge state distributions for single-layer and double-layer targets in Figs. \ref{fig:csd}(b) and \ref{fig:csd}(c), it is inferred that the maximum acceleration fields in the two cases should be close to each other. Note that the simulation results (red lines, t = 66 fs) in Figs. \ref{fig:pic_acceleration}(c) and \ref{fig:pic_acceleration}(d) also confirm this point. This implies that the charge state distributions of SHIs can be used to probe the ultraintense and transient electric fields, which are very difficult to be detected in experiments.  

It should be noticed that the highest charge state presented here is 61+ (Ar-like Au) is significantly lower than the recently reported value of 72+ in collisional ionization dominated laser-plasma interaction at the intensity of $ 3\times10^{21}\ {\rm W}/{\rm cm}^{2} $\cite{Hollinger2020np}. We believe this is because the field ionization overwhelms the collisional ionization for ultra-thin targets and ultrashort driving pulses at ultrahigh intensity in our case. The plasma collision rate\cite{huba1998nrl} can be estimated as $ \nu_{ee} \sim 10^{-6}n_{e}T_{e}^{-3/2}ln\Lambda_{ee} $, $ \nu_{ei} \sim 10^{-7}Z^{2}\mu^{-1/2}n_{i}T_{e}^{-3/2}ln\Lambda_{ei} $, and $ \nu_{ii} \sim 10^{-8}Z^{4}\mu^{-1/2}n_{i}T_{i}^{-3/2}ln\Lambda_{ii} $. Here $ \nu_{ee} $, $ \nu_{ei} $, $ \nu_{ii} $ is the electron-electron, electron-ion, ion-ion collision rate, respectively, $ \mu = m_{i}/m_{p} = 197 $, and $ ln\Lambda_{ee} $, $ ln\Lambda_{ei} $, $ ln\Lambda_{ii} $ are the Coulomb logarithms. In the simulation, we found that the plasma temperature $ T_{i} $ and $ T_{e} $ in the focal spot quickly rise up to more than 1 MeV after the on-target intensity is higher than $ 3\times10^{19}\ {\rm W}/{\rm cm}^{2} $. It can be calculated that $ \nu_{ee} < 10^{-5}\ {\rm fs}^{-1} $, $ \nu_{ei} < 10^{-3}\ {\rm fs}^{-1} $, $ \nu_{ii} < 10^{-3}\ {\rm fs}^{-1} $ by plugging in the simulation results. Within the interaction timescale of $ \sim $100 fs, the collision is insignificant in modifying the plasma dynamics.

\begin{figure}[b]
	\includegraphics{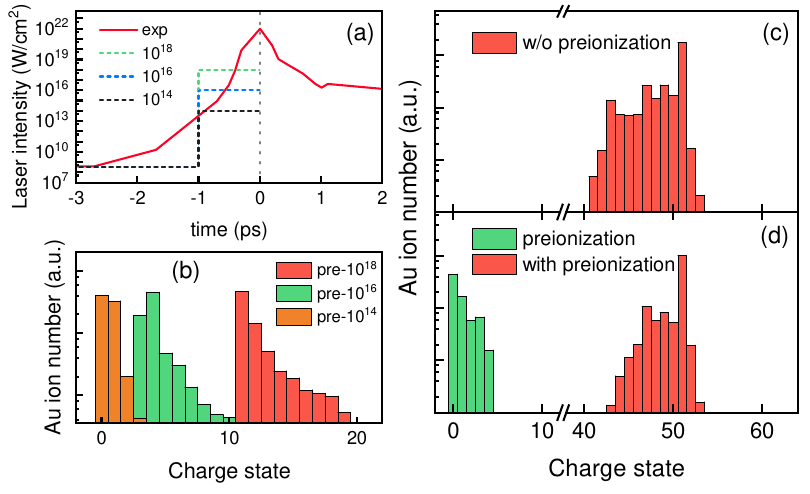}
	\caption{\label{fig:pre_csd} (a) Temporal profile of the laser pulses after the double plasma mirrors (red line), and the assumed 1-ps pedestal with different intensity (dashed lines). (b) Charge state distribution before the arrival of the main pulse. (c) Final charge state distribution without pre-ionization. (d) Charge state distribution before (green bars) and after the main pulse considering the pre-ionization cause by the rising edge of the laser pulse (red bars).}
\end{figure}

The collisional ionization may be important if the preceding laser intensity before the main pulse is high and lasts for a long time. We performed collisional 1D PIC simulations (up to -100 fs) for a 150 nm Au target to examine the pre-ionizaiton caused by the rising edge of the laser. The temporal profile\cite{Choi2020ol} of the input pulse measured by a third-order crosscorrelator is shown in Fig.~\ref{fig:pre_csd}(a). As one can see, the use of double-plasma mirrors significantly enhance the contrast, resulting an intensity of $\rm 5\times10^{13}\ W/cm^2$ at 1 ps before the main pulse. Figure~\ref{fig:pre_csd}(d) (green bars) shows the charge state distribution of Au ions before the arrival of the main pulse. The maximum state is 4+ and the average state is 0.64+. To investigate the influence of the pre-ionization on the Au acceleration, we subsequently performed a 2D simulation with initial charge states obtained from the collisional 1D simulation. The final charge state distribution, presented in Fig.~\ref{fig:pre_csd}(d) (red bars), has insignificant difference as compared to that without pre-ionization, as shown in Fig.~\ref{fig:pre_csd}(c), which verifies the opinion that collisionless ionization is dominant in our case. For the integrity of the study, we also performed 1D collsional simulations with 1ps pedestal profile at intensity of $10^{14}$, $10^{16}$, and $\rm 10^{18}\ W/cm^2$, respectively. As shown in Fig.~\ref{fig:pre_csd}(b), the pre-ionization become significant for $\rm 10^{18}\ W/cm^2$. But such a bad contrast is inapplicable for ion acceleration as well known.

\section{CONCLUSION AND OUTLOOK}

In conclusion, we successfully realize the generation of deeply ionized Au ions up to 1.2 GeV by using ultrashort femtosecond laser pulses at ultrahigh intensity. The charge state distribution measurement confirmed by 2D PIC simulations indicates the acceleration fields in near-critical-density double-layer targets has similar maximum strengths compared to those of single-layer targets, but decays much slower, which eventually leads to higher Au energies. The PIC simulations also reveal how the transverse position of Au ions in the focal spot influences their charge states and energy when collisionless ionization is dominant. Unlike low-Z ions and mid-Z ions, the maximum energy of the Au ions predominantly relies on the laser intensity over the pulse duration because of the difficulty in atomic ionization.

This work provides the first experimental results of SHIs acceleration at the intensity over $ 10^{22}\ {\rm W}/{\rm cm}^{2} $. The generation of 10s MeV/nucleon SHIs would be realistically expected by following our scheme at higher intensity. It paves the way to the applications requiring SHI beams with multiple charge states and high density such as injectors for heavy ion accelerator, heavy ion fission-fusion, and the generation of warm dense matter.

The presented self-calibrated diagnosis can be used not only to calibrate other heavy ions, but also to obtain the absolute charge state distributions of SHIs in laser-plasma interaction. The measured charge state distributions can serve as an additional tool to inspect the ionization and acceleration process. Such a method can be used as a universal diagnostic in laser-plasma experiments by doping Au or other SHIs on the surface of arbitrary targets.

\begin{acknowledgments}
The authors acknowledge the expertise of the Center for Relativistic Laser Science (CoReLS) staff. The work was supported by the Institute for Basic Science, Korea under the project code, IBSR012-D1, NSFC innovation group project (11921006), National Grand Instrument Project(2019YFF01014402), Natural Science Foundation of China (Grant Nos. 11775010, 11535001, 61631001) and Beijing outstanding young scientist Program. The PIC code EPOCH was in part funded by the United Kingdom EPSRC Grants No. EP/G054950/1, No. EP/G056803/1, No. EP/G055165/1, and No. EP/M022463/1. The simulations are supported by High-performance Computing Platform of Peking University.
\end{acknowledgments}

\appendix

\nocite{*}

\bibliography{manuscript}

\end{document}